\documentclass[pss]{wiley2sp} 
\usepackage{amsmath}
\usepackage{bm}              

\renewcommand{\epsilon}{\varepsilon}

\newcommand{\integral}[3]{\!\int\limits_{#2}^{#3}\!\!{\rm d}#1\;}

\newcommand{\expval}[2]{ \langle  #1 #2\ \!\! \rangle}

\newcommand{\elcre}[2]{ c^{\dagger}_{#1,#2}}
\newcommand{\elann}[2]{ c_{#1,#2}}

\newcommand{\e}{\mathrm e}

\newcommand{\vk}{{\bm k}}

\newcommand{\Imag}{\mathrm{Im}}

\newcommand{\hc}{\mathrm{h.c.}}

\begin{document}

\title{Quantum phase transition between antiferromagnetic and charge order in
  the Hubbard-Holstein model}

\titlerunning{Quantum phase transition between AFM and CO in the HH model }

\author{%
  Johannes Bauer\textsuperscript{\Ast,\textsf{\bfseries 1}},
  Alex C. Hewson\textsuperscript{\textsf{\bfseries 2}}}


\authorrunning{J. Bauer and A.C. Hewson}

\mail{
  \textsf{j.bauer@fkf.mpg.de}, Phone:
  +49-711-6891537}

\institute{%
  \textsuperscript{1}\,Max-Planck Institute for Solid State Research, Heisenbergstr.1,
  70569 Stuttgart, Germany\\
  \textsuperscript{2}\,Department of Mathematics, Imperial College, London SW7 2AZ,
  United Kingdom\\}

\received{XXXX, revised XXXX, accepted XXXX} 
\published{XXXX} 

\pacs{71.20.Tx,71.30.+h,71.38.-k,71.45.Lr,75.30.Fv,73.43.Nq} 

\abstract{%
%
%
%
    We explore the quantum phase transitions between two ordered states in the
    infinite dimensional Hubbard-Holstein model at half filling. Our study is
    based on the dynamical mean field theory (DMFT) combined with the
    numerical renormalization group (NRG), which allows us to handle both
    strong electron-electron and strong electron-phonon interactions. 
    The transition line is characterized by an effective electron-electron
    interaction. Depending on this effective interaction and the phonon
    frequency $\omega_0$ one finds either a continuous transition or discontinuous
    transition. Here, the analysis focuses on the behavior of the system
    when the electron-electron repulsion $U$ and the phonon-mediated
    attraction $\lambda$ are equal. We first discuss the adiabatic and antiadiabatic
    limiting cases. For finite $\omega_0$ we study the differences between the
    antiferromagnetic (AFM) and charge order, and find that when present the
    AFM state has a lower energy on the line.
%
%
}
\maketitle   

\section{Introduction}
 A feature of  strongly correlated systems is the existence of competing
 interactions on low energy scales which can lead to different types of
 symmetry breaking and different ground states. There can be transitions to
 various forms of  magnetic  order, to superconducting or charge ordered
 states; there also may be transitions between these states, and in some cases
 they even coexist.  Here we study  the competing effects arising from the
 inter-electron interactions and the electronic coupling to lattice modes, as
 described by the Hubbard-Holstein (HH) model,  and the competition between
 antiferromagnetic (AFM) and charge order (CO) at half filling. Our study is
 based on the dynamical mean field theory (DMFT) combined with the numerical
 renormalization group \cite{Wil75,BCP08} (NRG). The DMFT becomes exact in the
 limit of infinite dimensions \cite{GKKR96}, and it can generate
 non-perturbative solutions, such that electron-electron and electron-phonon
 interactions with arbitrary coupling strengths can be studied. 

There have been several applications of the DMFT-NRG method to study phase
transitions in the Hubbard-Holstein model \cite{BZ98,JPHLC04,KMH04,KHE05,SCCG05}.
There are various possible transitions to states of broken symmetry in this
model;  bipolaronic  (BP), CO, AFM and the superconducting (SC) state.
We restrict our attention here to the case of half-filling and zero temperature.
The transition first studied by the DMFT-NRG
method for this model did not include the possibility of either CO
or AFM \cite{MHB02,KMH04}.  There is, however, a
metal-insulator transition from the normal (N) to the BP state \cite{KMH04}.
If the possibility of transitions to CO and AFM are
included, then it is found that as the attractive term induced by
$\lambda=2g^2/\omega_0$ overcomes the repulsion due to $U$ the system changes
the ground state from AFM to CO \cite{Bau09pre,BH09pre}.
Close inspection shows that the transition between the CO and the AFM state 
generally does not exactly occur at $\lambda=U$, but at slightly larger values
of $\lambda$. Our results show evidence for a direct transition from an ordered
to an ordered state. There can be a continuous transition for smaller values of the
interactions $U$ and $g$ and discontinuous transitions for larger
interactions. 
The focus of this paper will be the behavior along the line $U=\lambda$.

\section{Model and method}
The Hamiltonian for the Hubbard-Holstein model is given by
\begin{eqnarray}
  \label{hubholham}
  H&=&-t\sum_{i,j,{\sigma}}(\elcre i{\sigma}\elann
j{\sigma}+\hc)+U\sum_in_{i,\uparrow}n_{i,\downarrow} \\
&&+\omega_0\sum_ib_i^{\dagger}b_i+g\sum_i(b_i+b_i^{\dagger})\Big(\sum_{\sigma}n_{i,\sigma}-1\Big).
\nonumber
\end{eqnarray}
$\elcre i{\sigma}$ creates an electron at lattice site $i$, and $b_i^{\dagger}$ a
phonon with oscillator frequency $\omega_0$, $n_{i,\sigma}=\elcre i{\sigma}\elann
i{\sigma}$. The electrons interact locally with strength $U$, and
their density is coupled to an optical phonon mode with coupling constant
$g$. We assume a semi-elliptic DOS, $\rho_0(\epsilon)=2\sqrt{D^2-\epsilon^2}/\pi
D^2$, with $D=2t$. $t=1$ sets the energy scale in the following.

For our calculations we assume a bipartite lattice with $A$ and $B$ sublattice, where
the matrix Green's function can be written in the form    
\begin{equation}
\underline{G}_{\vk,\sigma}(\omega) \!=\!
\frac1{\zeta_{A,\sigma}(\omega)\zeta_{B,\sigma}(\omega) -\epsilon_{\vk}^2}
\! \left(\!\!\!
\begin{array} {cc}
 \zeta_{B,\sigma}(\omega) & \epsilon_{\vk} \\
\epsilon_{\vk} & \zeta_{A,\sigma}(\omega)
\end{array}
\!\!\!\right),
\label{kgf}
\end{equation}
with $\zeta_{\alpha,\sigma}(\omega)=\omega+\mu_{\alpha,\sigma}-\Sigma_{\alpha,\sigma}(\omega)$,
$\alpha=A,B$, and $\vk$-independent self-energy.
For commensurate charge order we have $\mu_{A,\sigma}=\mu-h_c$, $\mu_{B,\sigma}=\mu+h_c$ and
$\Sigma_{B,\sigma}(\omega)=Un-\Sigma_{A,\sigma}(-\omega)^*$, with
$n=(n_A+n_B)/2$, $n_{\alpha}=\sum_{\sigma}n_{\alpha,\sigma}$. For the
AFM order one has $\mu_{A,\sigma}=\mu-\sigma h_s$, $\mu_{B,\sigma}=\mu+\sigma
h_s$, and the condition
$\Sigma_{B,\sigma}(\omega)=\Sigma_{A,-\sigma}(\omega)$.  We consider solutions
where the symmetry breaking fields vanish, $h_c,h_s\to 0$.
In the AFM case the $A$-sublattice magnetization,
$\Phi_{\rm afm}=m_A=(n_{A,\uparrow}-n_{A,\downarrow})/2$ serves as an order parameter.
For CO we define  $\Phi_{\rm co}=(n_A-1)/2$.

In the DMFT  this local  Green's function, and the self-energy
are identified with the corresponding quantities for 
an effective impurity model \cite{GKKR96}. One focuses for the calculations
on the properties of the $A$-sublattice. We can take the form of this
impurity model to correspond to an  Anderson-Holstein impurity model \cite{HM02} and
calculations are carried out as detailed, for instance in Ref. \cite{ZPB02,BH07c}.
We solve the effective impurity problem with NRG adapted to these cases with symmetry
breaking. For the logarithmic discretization parameter we take the value
$\Lambda=1.8$ and  keep about 1000 states at each iteration. The initial
bosonic Hilbert space is restricted to a maximum of 50 states.


\section{Behavior along $U=\lambda$}
As the electron-phonon coupling in  (\ref{hubholham}) is linear, the bosonic
field can be integrated out in a path integral framework, which yields an
effective electron-electron interaction of the form 
\begin{equation}
  U_{\rm eff}(\omega)=U+\frac{2g^2\omega_0}{\omega^2-\omega_0^2},
\label{Ueffom}
\end{equation}
The two terms are the competing interactions on different
energy scales. For large $\omega$, the electron repulsion $U$ dominates. Near
$\omega=\omega_0$ the retarded effective attraction due to the phonon starts
to play a role and in the limit $\omega\to 0$ the expression tends to 
$U_{\rm   eff}=U-\lambda$. We are studying the case $U=\lambda$, so $U_{\rm
  eff}=0$  at low energy. For the general behavior on all energy scales,
finite $\omega_0$,  and arbitrary coupling strength no simple description is
available. First insight can be gained by considering limiting cases.  

In the antiadiabatic limit, $\omega_0\to \infty$, where
$\lambda=2g^2/\omega_0$ is kept fixed,  $U_{\rm  eff}(\omega)$ becomes
completely independent of $\omega$ and tends to $U_{\rm   eff}=U-\lambda$, so
that the model then becomes equivalent to a pure Hubbard model with $U=U_{\rm
  eff}$. The system is then not ordered for any value $U=\lambda$. For any
finite $U_{\rm eff}>0$ the system is AFM ordered \cite{Don91} and for $U_{\rm eff}<0$ in
the CO or SC state \cite{MRR90}. We have
therefore a continuous transition from an ordered to an ordered state. In
the DMFT-NRG calculations we find that the smaller $\omega_0$ is the larger
the order parameter near the transition becomes. Thus we conclude that this
transition scenario persists for weak coupling and finite $\omega_0$.

\begin{figure}[!thbp]
\centering
\includegraphics[width=0.45\textwidth]{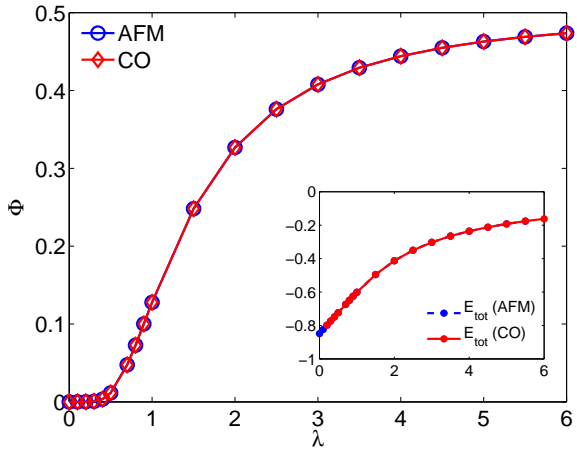}
\caption{(Color online) The mean field expectation values $\Phi$ along the
  $U=\lambda$-line. Inset: Total energy.}       
\label{mfphi}
\end{figure}

Another limit the adiabatic limit, $\omega_0\to 0$, can be studied transparently
in static mean field theory. The mean field self-energy reads,
\begin{equation}
\Sigma_{\alpha,\sigma}(\omega)=Un^{\alpha}_{-\sigma}-\lambda(n^{\alpha}-1),  
\end{equation}
independent of $\omega$. In order to determine
$n^{\alpha}_{\sigma}\equiv  \expval{\hat n_{\sigma}^{\alpha}}{}$,
we use the self-consistency equation
\begin{equation}
  n^{\alpha}_{\sigma}=\sum_{m=\pm}\integral{\epsilon}{}{}
\frac{\rho_0(\epsilon)u_{m,\sigma}^{\alpha}(\epsilon)}{1+\e^{\beta\omega_{m,\sigma}(\epsilon)}}.
\end{equation}
We have used the spectral function of the first element of the matrix Green's function for
the bipartite lattice in the form (\ref{kgf}),
\begin{equation}
  \rho_{\alpha,\vk,\sigma}(\omega)=-\Imag\frac{\zeta_{\bar \alpha,\sigma}(\omega^+)/\pi}
{\zeta_{A,\sigma}(\omega^+)\zeta_{B,\sigma}(\omega^+)-\epsilon_{\vk}^2},
\end{equation}
with $\omega^+=\omega+i\eta$.
We have introduced
\begin{equation}
\omega_{\pm,\sigma}(\epsilon_{\vk})=\frac{\Sigma_{A,\sigma}-\mu_{A,\sigma}+\Sigma_{B,\sigma}-\mu_{B,\sigma}}{2}
\pm E_{\sigma}(\epsilon_{\vk}),
\end{equation}
where
\begin{equation}
E_{\sigma}(\epsilon_{\vk})=\sqrt{\epsilon_{\vk}^2+
\frac{[\mu_{A,\sigma}-\Sigma_{A,\sigma}-(\mu_{B,\sigma}-\Sigma_{B,\sigma})]^2}4}.
\end{equation}
and
\begin{equation}
  u_{m,\sigma}^{\alpha}(\epsilon_{\vk})=\frac{\zeta_{\bar\alpha,\sigma}(\omega_{m,\sigma}(\epsilon_{\vk}))^2}
{\epsilon_{\vk}^2+\zeta_{\bar\alpha,\sigma}(\omega_{m,\sigma}(\epsilon_{\vk}))^2}.   
\end{equation}
The expression for the total energy reads at half filling
\begin{equation}
  E_{\rm mf}=
 E_{\rm kin}^{\rm mf}
+(U- 2\lambda)\Phi_{\rm co}^2-U\Phi_{\rm afm}^2+\frac U4,
\label{Emfsimp}
\end{equation}
where $E_{\rm kin}^{\rm mf}$ is the kinetic energy term \cite{BH09pre}.
We study exclusive order here, i.e., either CO or AFM; coexistence is
energetically not favorable.
We can see that for order parameters of equal magnitude the CO state has lower
energy for $\lambda>U$ and the AFM state otherwise. 
For $U=\lambda$ both states are degenerate. When we increase the interaction
$U=\lambda$ from zero we find that the order increases as illustrated in
Fig. \ref{mfphi}.   

\noindent
We see the typical mean-field exponential increase. One can, however, note a
suppression due to the competing interaction present. The total energy is shown
as an inset. 
The static mean field
solution in the adiabatic limit gives always a discontinuous transition
from CO to AFM state when $U$ or $\lambda$ are changed. 

\begin{figure}[!htbp]
\centering
\includegraphics[width=0.45\textwidth]{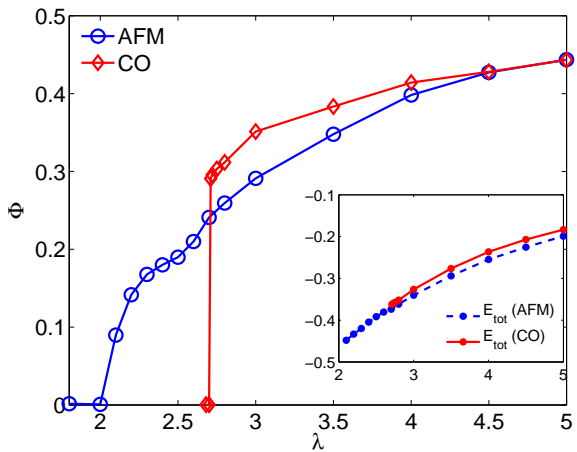}
\caption{(Color online) The expectation values $\Phi$ along the
  $U=\lambda$-line for $\omega_0=0.2$. Inset: Total energy.}       
\label{phi_trans_wo0.2}
\end{figure}

We now turn to the situation for finite $\omega_0$. This situation is accessed
with DMFT-NRG. In Fig. \ref{phi_trans_wo0.2} we show the
result for $\omega_0=0.2$. 
\noindent
For small values of $U=\lambda$ both order parameters are zero. From a certain
value of the interaction strength on solutions with finite AFM order can be
found, and for even larger interactions 
also solutions with finite CO exist. The point in the phase diagram separating
first and second order transition lies close to this point where both order
parameters are finite, $\lambda_{\rm fs}\simeq 2.7$. It will be
shifted to larger values $\lambda_{\rm   fs}$ when $\omega_0$ increases.
Our numerical results
give a continuous rise for $\Phi_{\rm afm}$, and a sudden increase for $\Phi_{\rm co}$.

From the total energies (inset) we can see that the AFM state has lower
energy and is therefore favorable along the $U=\lambda$-line.  Calculations in
the normal state reveal renormalized quasiparticles ($z<1$) with a small repulsive
effective quasiparticle interaction, which is consistent with this
observation \cite{BH09pre}. Thus, there is an asymmetry of AFM and CO
state for finite $\omega_0$. 

In summary we have analyzed the competing interactions in the HH model and
focused on the situation where $U=\lambda$. For $\omega_0\to\infty$ there is no
order along this line in the $U-\lambda$-phase diagram, whereas for
$\omega_0\to 0$ and finite interactions the system is ordered with either AFM
or CO, which are degenerate. For intermediate values of $\omega_0$ there is
no order for small couplings and AFM or CO can be present for larger
couplings, where the AFM state is found to have lower energy. 

\begin{acknowledgement}
We wish to thank O. Gunnarsson, G. Sangiovanni, and R. Zeyher for
helpful discussions, and W. Koller and D. Meyer for their earlier
contributions to the development of the NRG programs.
\end{acknowledgement}

%
 \bibliographystyle{pss}
%

\providecommand{\WileyBibTextsc}{}
\let\textsc\WileyBibTextsc
\providecommand{\othercit}{}
\providecommand{\jr}[1]{#1}
\providecommand{\etal}{~et~al.}

\end{document}